\documentstyle[12pt]{article}
\begin{document}
\input{psfig}
\abovedisplayskip=4.37ex
\belowdisplayskip=4.37ex
\abovedisplayshortskip=4.37ex
\belowdisplayshortskip=4.37ex
\newcommand{\eee}{{\em et al.}}
\newcommand{\tot}{_{\rm tot}^{pp}}
\newcommand{\sof}{_soft}
\newcommand{\lin}{\lambda_{\rm in}(E)}
\newcommand{\lob}{\lambda_{obs}(E)}
\newcommand{\pair}{ _{\rm in}^{p- \rm air}}
\newcommand{\ppp}{\hspace{3em}}
\newcommand{\igual}{\: = \:}
\newcommand{\meio}{\frac{1}{2}}
\newcommand{\nnn}{\noindent}
\newcommand{\sigin}{\sigma _{\rm in}^{p- \rm air}}
\newcommand{\sigab}{\sigma _{abs}^{p- \rm air}}
\newcommand{\sitot}{\sigma _{\rm tot}^{pp}}
\baselineskip=24pt
\title{ \huge \bf
Diffractive Contribution to the Elasticity
and the Nucleonic Flux in the Atmosphere}
\author{\large
J. Bellandi,
R. J. M. Covolan, \underline{A. L. Godoi} and J. Montanha \\ \\
Instituto de F\'{\i}sica {\em Gleb Wataghin} \\
Universidade Estadual de Campinas, Unicamp \\
13083-970 \  Campinas \  SP \  Brazil }
\date{}
\maketitle
{\large E-mail address: adriana@ifi.unicamp.br}

PACS NUMBERS: 13.85.Tp, 96.40.De

\baselineskip=24pt
\vspace{1cm}

\begin{abstract}
\baselineskip=24pt
We calculate the average elasticity
considering non-diffractive and single diffractive
interactions and perform an analysis of the cosmic-ray
flux by means of an
analytical solution for the nucleonic
diffusion equation. We show that the diffractive contribution
is important for the adequate description of the nucleonic
and hadronic fluxes in the atmosphere.
\end{abstract}

\newpage

It is well known that the evolution of the nucleonic
cosmic ray component is controlled by two physical
quantities related to high energy hadron interactions:
the interaction mean-free-path $\lambda_{\rm in}^{p- \rm air}$, which is
inversely proportional to the inelastic proton-air
cross section, $\sigin$, and
the average elasticity $\langle x \rangle ^{p- \rm air}$, the fraction
of energy retained by the incident particle after a collision.
It was shown in Refs. \cite{bel579,bel149} that, when
one supposes the
interaction mean-free-path and the mean
elasticity as energy-dependent
quantitites, the analytical solution for
the nucleonic diffusion equation in the atmosphere is
given by
\begin{equation}
F_{N}(E,{\tt t}) \igual
N_{0}E^{-(\gamma + 1)}
\exp
\left[
-
\frac{{\tt t}(1 - (\langle x \rangle ^{p- \rm air}) ^{\gamma})}
{\lambda_{\rm in}^{p- \rm air}(E)}
\right]
\end{equation}
where {\tt t} is the atmospheric depth and
$N_{0}E^{-(\gamma + 1)}$ is the primary differential
spectrum.

In previous papers \cite{bel579,bel149,our} we have discussed the importance
of the energy dependence of the leading particle spectrum,
which was done through the mean inelasticity.
As a first approach, we have neglected
the diffractive interactions in
the proton-air collisions.
By diffractive interactions we mean processes like
\[ a \: + \: b \rightarrow a \: + X \]
where particle $b$ is excited to a system X with the
same quantum numbers and characterized by an invariant mass $M$.
This kind of process is called {\em single} diffraction
(for details see, for instance, Ref. \cite{mat}).
In general, diffractive interactions are neglected in
cosmic-ray physics (mostly in analytical calculation of cascades),
because their contributions are {\em a priori} considered
to be very small.

In this paper, we consider that the leading particle
distribution has two contributions, namely non-diffractive
(ND)
and single-diffractive (SD), and we show
that, although the latter is really small, its effect is
relevant for an accurate description of cosmic-ray fluxes.

In spite of its simplicity, Eq. (1) has two parameters, $\langle x\rangle
^{p- \rm air}$ and $\lambda_{\rm in} ^{p- \rm air}$, which must
contain all dynamic aspects of the hadron collisions occurring
in the atmosphere. Therefore, in the following we shall describe how these
aspects can be taken into account and, mainly, how to compute
diffractive effects in the leading particle distribution.

In Eq. (1), the interaction mean-free-path is given by
\begin{equation}
\lambda_{\rm in}^{p- \rm air}(E) \igual
\frac{2.4 \times 10^{4}}{\sigin \, (\rm mb)} \: \:
({\rm g/cm}^{2}),
\end{equation}
with the $p$-air inelastic cross section
calculated here by means of the Glauber model \cite{glau}
\begin{equation}
\sigin \igual
\int d^{2}b
\left\{
1 - \exp
\left[
-\sigma \tot T(b)
\right]
\right\},
\label{glau}
\end{equation}
\nnn
where $b$ is the impact parameter and
$T(b)$ is the nuclear thickness
\begin{equation}
T({\bf b}) \igual \int_{-\infty}^{+\infty}
\rho({\bf b},z)dz
\end{equation}
given in terms of the
nuclear distribution $\rho({\bf b},z)$ (see Ref. \cite{our}).
For $\sitot$, we use in Eq. (3) the best fit of UA4/2
Collaboration \cite{ua4}.

Now we need to establish $\langle x \rangle ^{p- \rm air}$
in order to calculate the nucleonic flux.
We begin by writting the partial average elasticities, SD
and ND, as
\begin{equation}
\langle x \rangle_{\rm SD}
\igual
\frac{ \int_{0}^{1}
x \frac{ d \sigma_{\rm SD}} {dx} dx }
{\sigma_{\rm SD}}
; \hspace{1cm}
\langle x \rangle_{\rm ND}
\igual
\frac{ \int_{0}^{1}
x \frac{ d \sigma_{\rm ND}} {dx} dx }
{\sigma_{\rm ND}},
\end{equation}
which are correctly normalized to compound the average elasticity
in $pp$ collisions by
\begin{equation}
\langle x\rangle ^{pp} \igual
\frac{\sigma_{\rm SD}^{pp}}{\sigma_{\rm in}^{pp}}
\langle x\rangle _{\rm SD}^{pp} \: + \:
\frac{(\sigma_{\rm in}^{pp} \: - \: \sigma_{\rm SD}^{pp})}
{\sigma_{\rm in}^{pp}}
\langle x\rangle _{\rm ND}^{pp}.
\end{equation}

In the above expression, we are assuming that
$\sigma_{\rm in}^{pp} \: = \:
\sigma_{\rm SD}^{pp} \: + \: \sigma_{\rm ND}^{pp}$, where
$\sigma_{\rm in}^{pp}$
is given by the Landshoff parametrization,
$\sigma_{\rm in}^{pp} \igual 56 s^{-0.56} + 18.16 s^{0.08}$
\cite{land}.

In the whole calculation, whose results we shall show farther on,
we have used two models from which we borrowed the
ND and SD distributions.
For the non-diffractive elasticity
($\langle x \rangle _{\rm ND}^{pp}$), we use the {\em Interacting
Gluon Model (IGM)}, revised by Dur\~aes \eee \cite{dunga} with
the purpose of including semi-hard interactions, responsible
for  mini-jets events.
In Ref. \cite{our}, we have  shown
that the correspondent non-diffractive leading
particle distribution produces reasonable
results for  $pp$ total cross section and
for  $p$-air inelastic cross section.

In order to include the diffractive contribution, we use
the leading particle distribution of the Covolan-Montanha
model \cite{mon}, which reads
\begin{equation}
\frac{d \sigma_{\rm SD}}
{dx} \igual
\int
\frac{
d^{2} \sigma_{\rm SD}}
{dtdM^{2}} dt
\end{equation}
where, for $pp$ interactions, the invariant cross section is given by
\begin{eqnarray}
\frac{ d^{2} \sigma_{\rm SD}}
{dtdM^{2}} & = &
\frac{
(3 \beta_{p} G_{p}(t) )^{2}}{16 \pi}
s^{2 \alpha_{\bf I\! P}(t) - 1}
\sigma_{{ \bf I\! P}\,p}(M^{2}), \\
&& \nonumber \\
\sigma_{{\bf I\! P}\,p}(M^{2}) & = & 3\beta_{p}\,\xi\, \langle
r^{2}_{p}(M^{2}) \rangle,
\label{mon}
\end{eqnarray}
with $\beta_{p} = 2.502 \, {\rm GeV}^{-1}, \: \xi = 0.0764 \, {\rm GeV}$,
$ G_{p}(t)$ is the electric form factor of the proton,
$\alpha_{\bf I\! P}(t) = 1.08 + 0.25t$, where
$t$ is the squared four-momentum transfer, and
$ \langle r^{2}_{p}(M^{2}) \rangle = 12.75+0.84\ln M^{2}$.
We remind the reader that, in  diffractive processes, we have
$x \igual 1 - M^2/s$.

As we intend to calculate hadronic interactions in the atmosphere,
$\langle x \rangle ^{pp}$ must be corrected to include
air effect. This is done by the procedure given in Ref. \cite{aza},
\begin{equation}
\langle x \rangle ^{p- \rm air}_{c}
\igual \sum_{n=1}^{n_{\rm max}}
P_{n} ( \langle x \rangle ^{pp})^{n},
\end{equation}
\vspace{-0.5cm}
\nnn where
\vspace{-0.5cm}
\begin{equation}
P_{n} \igual
\frac{
\int d^{2}bP_{n}(b)}
{\sigma \pair}
\end{equation}
\vspace{-0.5cm}
\nnn and
\vspace{-0.5cm}
\begin{equation}
P_{n}(b) \igual
\frac{1}{n!}
\left[
\sigma \tot T(b)
\right]^{n}
\exp
\left[
-\sigma \tot
T(b)
\right].
\end{equation}

Here $P_{n}$ is the probability of n-fold
collisions of the primary nucleon
inside the nucleus and $n_{\rm max}$, the maximum number
of collisions, is roughly given by $2.3A^{1/3}$ \cite{aza}.

The diffractive contribution included in Eq. (6) comes from inelastic
interactions among the incident hadron and the nucleons
inside the struck nucleus. In addition to this contribution, it is
necessary to consider the diffractive dissociation of the nucleus
as a whole. For this reason, we add to $<x>^{p- \rm air}_{c}$ (once again,
in a weighted way) a second component coming from nuclear diffractive
processes, so that our final expression is
\begin{equation}
\langle x\rangle ^{p- \rm air} \igual
\frac{\sigma_{\rm SD}^{p- \rm air}}{\sigin}
\langle x\rangle _{\rm SD}^{p- \rm air} \: + \:
\frac{(\sigin \: - \: \sigma_{\rm SD}^{p- \rm air})}
{\sigin}
\langle x\rangle ^{p- \rm air}_{c}.
\end{equation}

In order to calculate $\langle x\rangle_{\rm SD}^{p- \rm air}$ and $\sigma_{\rm
SD}
^{p- \rm air}$, we use an extension of the Covolan-Montanha model to
nuclear diffractive interactions which is made by a radial scaling
\cite{mon}, {\em i.e.} multiplying
Eq. (\ref{mon}) by
\begin{equation}
\frac{\langle
r^{2}_{N}(A)\rangle^{\frac{1}{2}}}{\langle
r^{2}_{p}\rangle^{\frac{1}{2}}},
\end{equation}
where
$\langle r^{2}_{N}(A) \rangle ^{\frac{1}{2}}
\igual
1.096 \, A^{\frac{1}{3}}
- 0.41 \, A^{-\frac{1}{3}}$ \cite{bel149}
corresponds to the atomic radius as a
function of the atomic mass $A$, and
$\langle  r^{2}_{p} \rangle ^{\frac{1}{2}}
\igual
0.197\sqrt{12.75+0.84\ln s}$
is the hadronic radius of the proton, both
given in fermis.

As experimental data for the nucleonic flux cover
a large range of energy, $1 < E \, ({\rm GeV}) < 10^{3}$,
we need to stablish a
{\em cutoff} for our elasticity at
laboratory energy $E=50 \, {\rm GeV}$; for lower
energies, the  $\langle x \rangle ^{pp}$ is kept constant
and the $\langle x \rangle _{\rm SD}$ is turned off. This is
necessary basically because of two reasons: 1) for energies below
50 ${\rm GeV}$,
the elasticity given by the IGM drops steeply to very low values;
2) at these low energies the Covolan-Montanha model, which is based
on the Triple-Pomeron formalism, is out of its validity range and
does not hold.
In order to avoid a discontinuity of the calculated flux
due to this abrupt change in the elasticity, we use
its value at $50 \, {\rm GeV}$ as a constraint and calculate the
correspondent $\langle x \rangle ^{pp}$. Then, this value
($\langle x \rangle ^{pp}=0.6125$)
is kept constant for lower energies, only being corrected to
include air effect.

In the Fig. 1, the calculated nucleonic flux is
compared with experimental data measured at
sea level \cite{wolf,asht}, using the
Ryan primary spectrum \cite{ry},
$N_{0} \igual 2 ({\rm cm^{2}.s.sr.GeV})^{-1}$ and $\gamma \igual 1.5$.
With only
ND contribution,
our solution underestimates the nucleonic
flux, as shown by the dashed line in Fig. 1.
The final result, including both contributions
(ND and SD) and free of
parameters, is shown by the solid line.
One can see a significant improvement
of our theoretical calculation, when diffractive effects are
taken into account.

Recently, measurements of hadronic fluxes at sea level,
in the energy range $1 < E \, ({\rm GeV}) < 10^{5}$, were performed
by the hadronic calorimeter of the KASCADE experiment
\cite{mie}. In order to compare the hadronic
flux calculated by means of the formalism developed in this
paper with those data,  we borrow
the KASCADE parametrization for the pion to nucleon
ratio
\begin{equation}
 R \igual
\frac{
\pi^{+} + \pi^{-}}{n + p} \igual
0.04 \: + \: 0.27 \log (E/{\rm GeV}),
\end{equation}
and we correct with this factor our nucleonic flux
to obtain the hadronic flux
\begin{equation}
F_{H}(E,{\tt t}) \igual
(1 \: + \: 2R)F_{N}(E,{\tt t}).
\end{equation}

In the Fig. 2, the analytical solution of the hadronic flux
is compared with
KASCADE's experimental data, showing a good agreement
between them.

In conclusion,
the diffractive contribution to the mean elasticity
is taken into account by two mechanisms. The first one
comes from the inelastic interactions among the incident
hadron and the nucleons of the target, while the second
considers the diffractive dissociation of the nucleus
as a whole. The net effect of these two corrections is
to increase the value of the mean elasticity. This enables us
to perform a satisfactory description of the nucleonic
and hadronic fluxes at sea level, without further
parameters.

We would like to thank the Brazilian governmental
agencies CAPES, CNPq and FAPESP for financial
support.

\newpage

\nnn
{\Large \bf Figure Captions}
\vspace{0.8cm}

\nnn
FIGURE 1:
Nucleonic flux at sea level. Dashed line:
only ND contribution.
Solid line:
ND and SD contributions.
Open circles, Ref. [11]. Full circles, Ref. [12].
\vspace{1cm}

\nnn
FIGURE 2:
Hadronic flux at sea level.
The solid line is our calculated hadronic flux.
Stars, Ref. [14]. Open diamonds, Ref. [15].

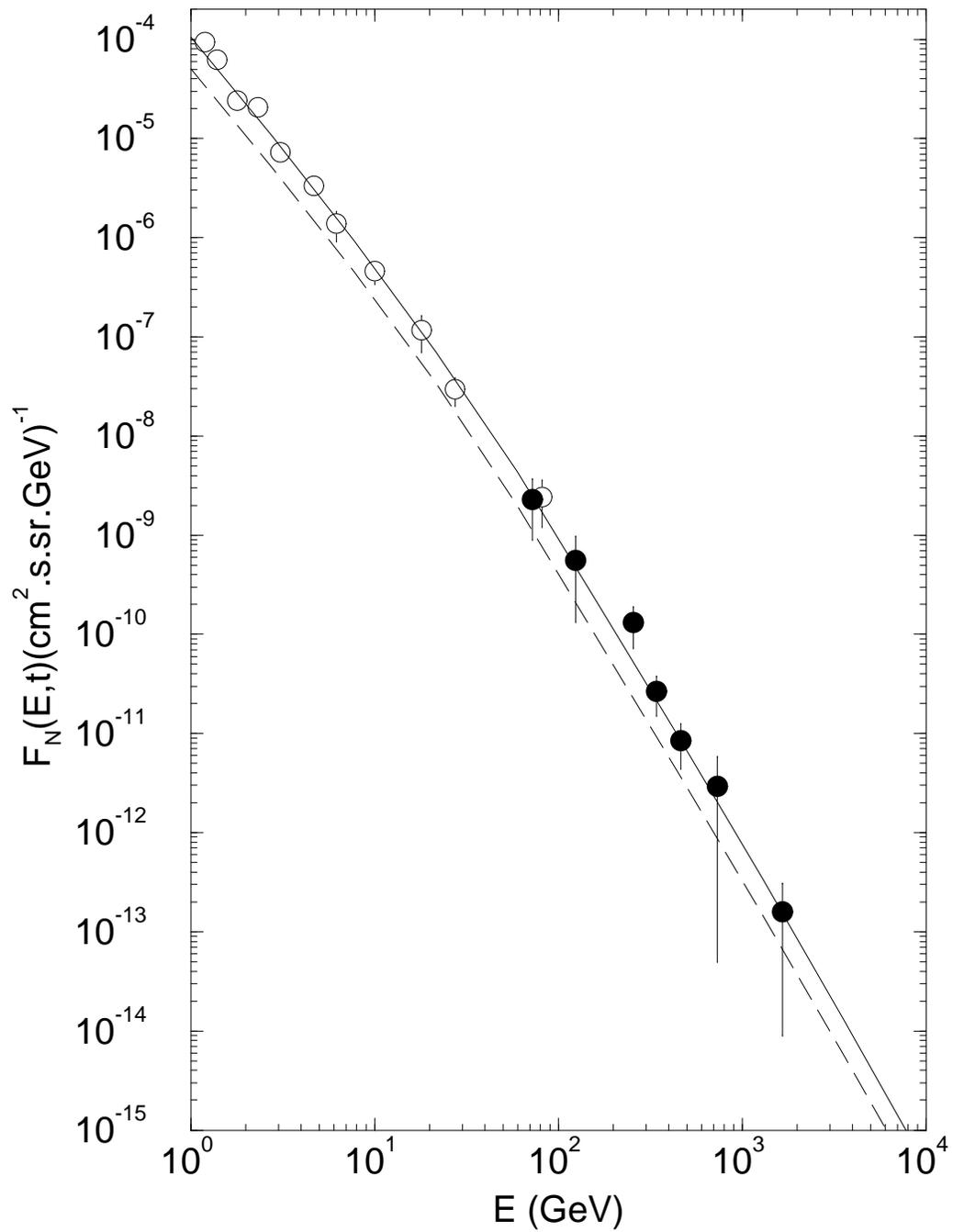
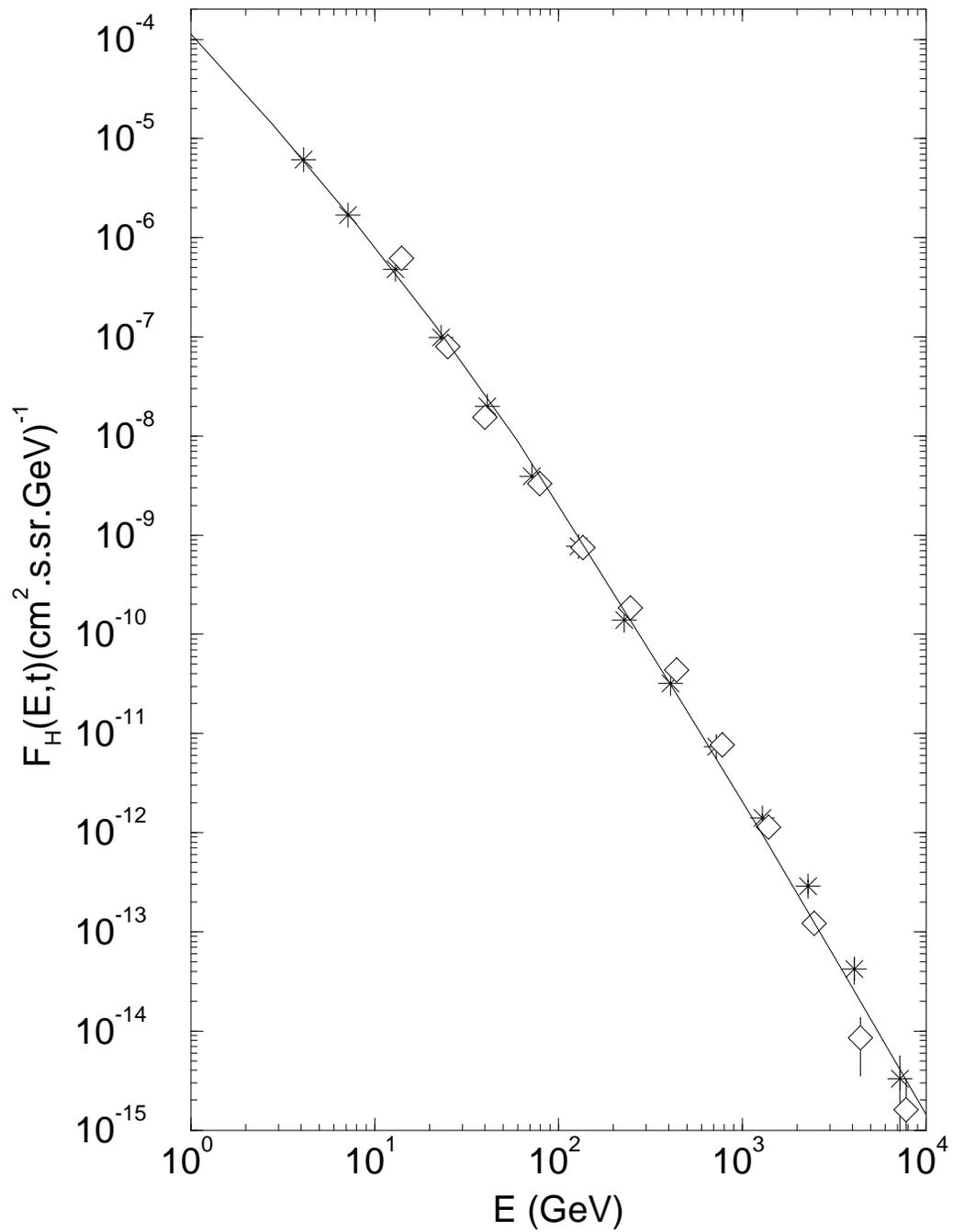
\end{document}